\documentclass[12pt]{article} 
\usepackage{amssymb}
\usepackage{graphicx}
\begin{document} 
\begin{center}
{\large \bf The interaction region of high energy protons }

\vspace{0.5cm}                   

{\bf I.M. Dremin}

\vspace{0.5cm}                       

         Lebedev Physical Institute of RAS, 

\medskip

        National Research Nuclear University "MEPhI"     

\bigskip

{\large \bf Content}

\end{center}

{\bf 1. Introduction

2. Main facts and relations

3. The geometry of the interaction region

4. New tendencies of inelastic interactions

5. Elastic scattering outside the diffraction cone

6. Conclusions

References}

\begin{abstract}

New experimental data about proton-proton collisions obtained at the LHC 
allow to widen strongly the energy interval where one gets some knowledge
about the structure of their interaction region.

Using the unitarity relation in combination with experimental data about 
the elastic scattering in the diffraction cone, we show how the shape 
and the darkness of the inelastic interaction region of colliding protons 
change with increase of their energies. In particular, the collisions become 
fully absorptive at small impact parameters at LHC energies that results in 
some special features of inelastic processes. Possible evolution of the shape 
from the dark core at the LHC to the fully transparent one at higher energies 
is discussed that implies the terminology of the black disk 
would be replaced by the black toroid. Another regime could appear with the
approach to asymptotics.

The parameter which determines the opacity of central collisions also plays a
crucial role in the behavior of the differential cross section of elastic
scattering outside the diffraction cone where the predictions of all 
phenomenological models failed at LHC energies. The role of the ratio of real 
to imaginary parts of the elastic scattering amplitude at non-forward 
scattering becomes decisive there as seen from the unitarity condition. 
The obtained results allow to estimate this ratio outside the diffraction cone 
for the first time by comparison with experiment at LHC energies, and it happens
to be drastically different from its values measured at forward scattering. 
Moreover, the behaviours of the real and imaginary parts separately
differ in different phenomenological models and in the approach based on the 
unitarity condition. This problem is still waiting for its resolution.

All the conclusions are only obtained in the framework of the indubitable 
unitarity condition using experimental data about the elastic scattering of 
protons in the diffraction cone without references to quantum chromodynamics 
(QCD) or phenomenological approaches. 
\end{abstract}

\section{Introduction}

In this paper, I concentrate on two problems which became of topical interest 
nowadays in connection to experiments on particle interactions done at the 
highest available energies of the Large Hadron Collider (LHC) in CERN 
(Switzerland). To be more definite, our knowledge of the shape and opacity of 
the interaction region of two colliding protons and the behavior of their 
elastic scattering amplitude at various transferred momenta will be discussed. 

I try to present them in a way most adoptable for newcomers.

The general common approach to both problems considering the irrefutable statement 
that the total probability of all possible processes must be equal to 1 will be
used. It is called the unitarity condition and will be applied to information 
stemming from experiment about the elastic scattering of protons at small
angles within the diffraction cone for the first problem and at larger angles
outside it for the second problem. The generality of the approach guarantees
the certainty of the obtained results. At the same time, surely, it can not 
substitute the knowledge of the dynamics of the process but helps get some 
interesting conclusions about the problems to be approached. That is especially
important in view of the limited applications of QCD
to quantitative description of experimental data. In addition, some results of 
the phenomenological models are briefly discussed and confronted to our 
conclusions as well. The usage in the present paper {\bf only} these two 
indubitable ingredients - the unitarity condition and experimental results about 
the elastic scattering - is decisive for the confidence in derived conclusions.

Why are these problems so important?

The knowledge of the elastic scattering amplitude in the wide range of energies 
and scattering angles would provide some guides for  
QCD which is practically unapplied yet to this process. Several attempts to 
use diagrams of elastic scattering of hadrons containing incoherent quarks and 
gluons were done only in the asymptotical freedom regime of QCD at large 
transverse momenta with some phenomenological arguments added. Probably, 
the small angle scattering would ask for account of coherent 
states of partons in initial protons. The lack of this knowledge prevents
further progress in this field. Any guesses from experiment about the behavior 
of real and imaginary parts of the amplitude are very desirable. 

The parton content of hadrons and their spatial interaction region help  
visualize the collision processes and compare them at different energies.  
The decisive role is here played by our knowledge of elastic scattering at
rather small angles in the diffraction cone. Herefrom we learn about the 
special regime with the black central core of the interaction region of protons 
with the total energy in the center of mass system 7 TeV observed at the LHC.
In particular, this knowledge helps in developing some models of inelastic 
processes when their contribution to the unitarity relation is disentangled 
from elastic terms. Moreover, the further evolution of the spatial region with 
energy can be speculated so that it leads to some intriguing predictions.

Definite conclusions about transverse momentum dependence (but, unfortunately, 
not about energy dependence) of the elastic scattering amplitude
at larger angles were obtained from the unitarity condition albeit 
with some adjustable parameters. Fits of experimental data in this region ask
for strongly enlarged role of real part of the amplitude compared to the 
diffraction cone. What concerns phenomenological 
models, their first attempts to predict the outcome of LHC experiments outside 
the diffraction cone failed. Their special feature is zero of the imaginary 
part at the position of the dip of the differential cross section. No such
zero is required by the unitarity condition. Till now, there is no definite 
consensus of these approaches about the behavior of the elastic scattering 
amplitude there. 

\section{Main facts and relations}

Colliding high energy hadrons can either scatter elastically when only two of 
them appear at the final stage without changing their nature or produce
some new particles in inelastic processes. Kinematics of elastic scattering
is very simple. It is described by two variables: the squared total energy 
$s=4E^2$, where $E$ is the energy of one of partners in the center of mass 
system, and the four-momentum transfer 
squared $-t=2p^2(1-\cos \theta )$ with $\theta $ denoting the scattering angle 
and $p$ the momentum in the center of mass system. For inelastic processes
the kinematics is much more complicated. Therefore, to avoid some complications,
it is quite natural to try to get at the first stage some knowledge about the 
dynamics of the whole process approaching from
the analysis of elastic scattering and using such general relation as the
unitarity condition. It follows from the irrefutable statement that the
total probability of all (elastic + inelastic) processes should be equal to 1.
In this way it relates these two channels of the reaction albeit in rather 
average integrated form. It is the mainstream of the approach adopted in 
the present paper.
 
The experimental data about elastic scattering at a given energy are not 
very abundant. The only information about this process comes from the 
measurement of the differential cross section as a function of
the transferred momentum at the experimentally available values of $t$
and of the ratio of the real and imaginary parts of the elastic scattering 
amplitude $f(s,t)$ $\rho (s,t)={\rm Re}f(s,t)/{\rm Im}f(s,t)$ just in forward 
direction $t=0$ $\rho (s,0)=\rho _0$ but not at any other values of $t$. The latter one is obtained
from studies of interference between the nuclear and Coulomb contributions to 
the amplitude  $f$ which becomes practically noticable only in the
near-forward direction.  

The differential cross section is related to the 
scattering amplitude $f(s,t)$ in a following way
\begin{equation}
\frac {d\sigma }{dt}=\vert f(s,t)\vert ^2.
\label{dsdt}
\end{equation}
Thus, from measurements of the differential cross section at any energy 
of colliding particles we get the knowledge only about the modulus of the 
amplitude at the experimentally available values of $t$. The typical shapes 
of the differential cross section at small and larger values of $\vert t\vert $ 
are demonstrated in Fig. 1 for the LHC energy $\sqrt s$=7 TeV borrowed from
\cite{totem1, totem2}.
\begin{figure}
\includegraphics[width=7cm, height=7cm]{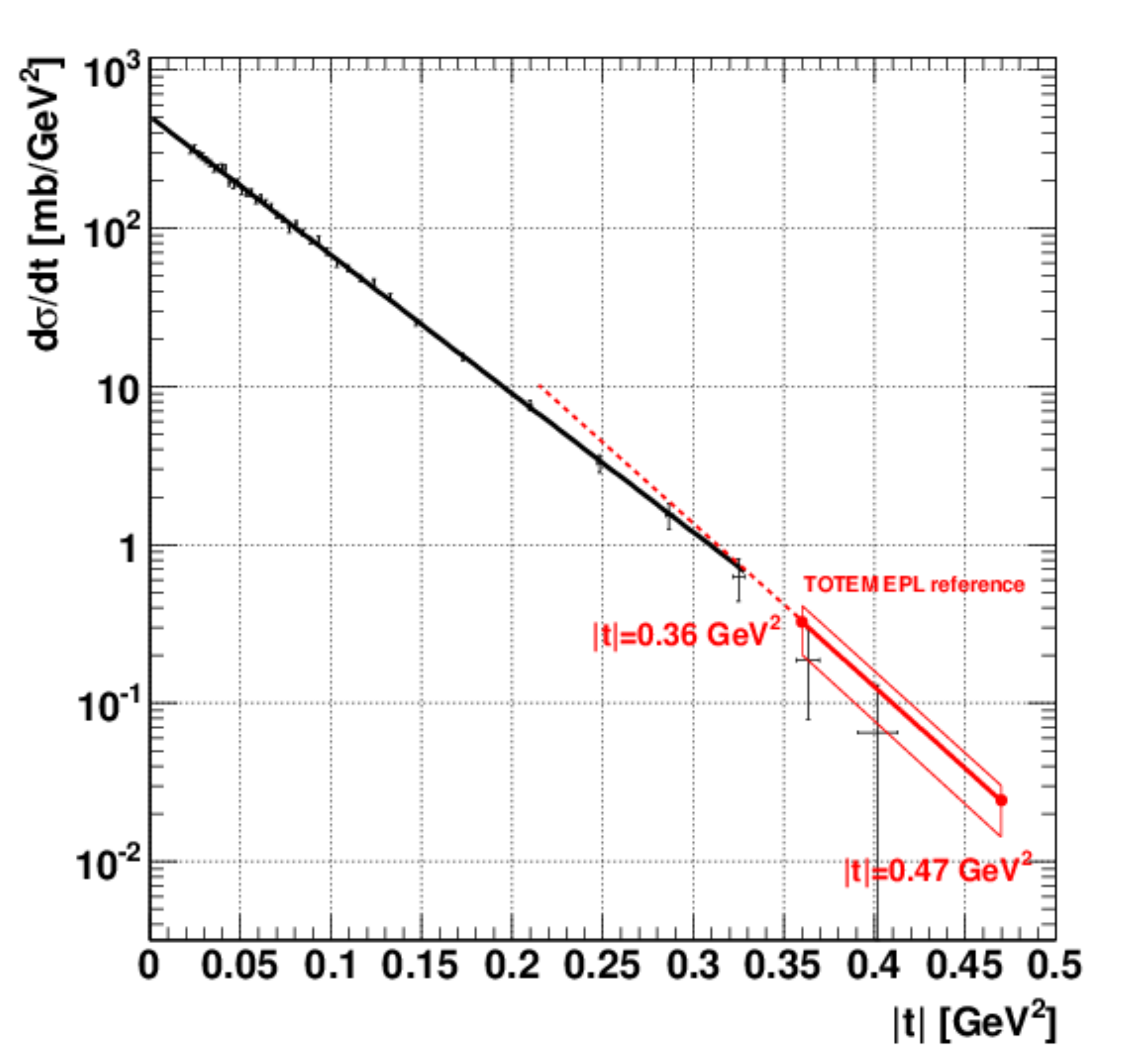} 
\includegraphics[width=7cm, height=7cm]{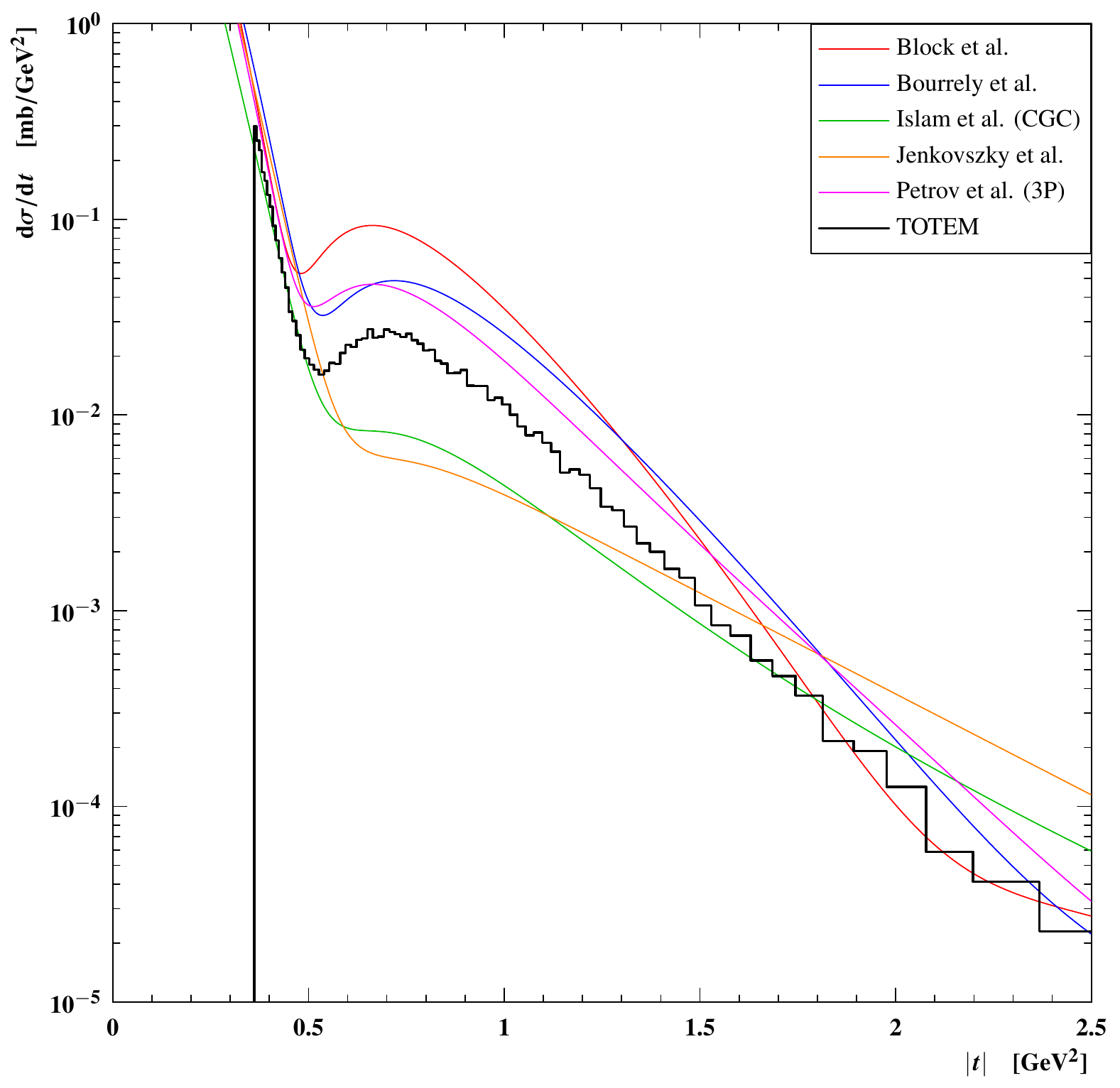}  

Fig. 1. The differential cross section of elastic proton-proton scattering at 
the energy $\sqrt s$=7 TeV measured by the TOTEM collaboration. \\
 Left: The region of the diffraction cone with the 
$\vert t\vert $-exponential decrease \cite{totem1}.

Right: The region beyond the diffraction peak \cite{totem2}. 
The predictions of five models are demonstrated.
\end{figure}

The most prominent feature of the plots in Fig. 1 is the fast decrease of the 
differential cross section with increasing transferred momentum
$\vert t\vert $. As a first approximation at present energies, it can be 
described at comparatively small transferred momenta in Fig. 1(left) 
by the exponential shape with the slope $B$ such that
\begin{equation}
\frac {d\sigma }{dt}=\frac {\sigma ^2_t}{16\pi }\exp (-B\vert t\vert ),
\label{expB}
\end{equation}
where $\sigma _t$ denotes the total cross section.
This region is called the diffraction peak. The peak becomes higher and its 
width shrinks with increasing energy because both the cross section and the 
slope increase with energy. The slope slightly depends on $t$ if more 
careful fits of experimental data are attempted as seen in Fig. 1(left). 
It diminishes somewhat at energies up to ISR with increasing transferred 
momentum while starts increasing at the LHC energy 7 TeV (see Fig. 1(left)).
Moreover, some oscillations around the exponent were found \cite{anti} at the 
energy $\sqrt s \approx 11$ GeV. The review of the early data can be found
in \cite{zrt}. Interesting by themselves, these details will be not very
important further for our approach because the integrals of (\ref{expB})
will be used where they can be accounted by slight variation of $B$.
In fact, the values of $B$ at small $\vert t\vert < 0.3$ GeV$^2$ are important.

At larger values of $\vert t\vert $ outside the peak we observe the dip and 
slower decrease of the plots with $\vert t\vert $ compared to the diffraction 
cone (see Fig. 1(right)).  Let us note that the normalization of the amplitude 
is fixed by Eqs (\ref{dsdt}), (\ref{expB}).

To disentangle the real and imaginary parts of the amplitude from the ratio
$\rho $, one needs the help from theorists.
From the theoretical side, the most reliable information comes from the 
unitarity condition. The unitarity of the 
$S$-matrix $SS^+$=1 imposes definite requirements on the amplitude of elastic
scattering $f(s,t)$ and amplitudes of inelastic processes $M_i$. 
In the $s$-channel it looks \cite{anddre1, ufnel} like 
\begin{eqnarray}
{\rm Im}f(p,\theta )= I_2(p,\theta )+g(p,\theta )= \nonumber  \\
\frac {s}{8\pi ^{3/2}}\int \int d\theta _1
d\theta _2\frac {\sin \theta _1\sin \theta _2f(p,\theta _1)f^*(p,\theta _2)}
{\sqrt {[\cos \theta -\cos (\theta _1+\theta _2)] 
[\cos (\theta _1 -\theta _2) -\cos \theta ]}}+g(p,\theta ).
\label{unit}
\end{eqnarray}
The region of integration in (\ref{unit}) is given by the conditions
\begin{equation}
\vert \theta _1 -\theta _2\vert\leq \theta ,       \;\;\;\;\;
\theta \leq \theta _1 +\theta _2 \leq 2\pi -\theta .
\label{integr}
\end{equation}
The non-linear integral term represents the two-particle intermediate states of the 
incoming particles. The function 
\begin{equation}
g(p,\theta )\propto \sum _i\int d\Phi _i M_iM_i^*(\theta )
\label{gmm}
\end{equation}
represents the shadowing contribution of the inelastic processes to the 
imaginary part of the elastic scattering amplitude. Following \cite{hove} it is 
called the overlap function. This terminology is ascribed to it because the 
integral there defines the overlap within the corresponding phase space 
$d\Phi _i$ between the matrix element $M_i$ of the $i$-th inelastic channel and 
its conjugated counterpart with the collision axis of initial particles turned 
by the angle $\theta $ of proton scattering in the elastic process. It is 
positive at $\theta =0$ but can change sign at $\theta \neq 0$ due to
the relative phases of inelastic matrix elements $M_i$'s.

At $t=0$ it leads to the optical theorem 
\begin{equation}
{\rm Im}f(s,0)=\sigma _t/4\sqrt {\pi}
\label{opt}
\end{equation}
and to the general statement that the total cross section is the sum of 
cross sections of elastic and inelastic processes
\begin{equation}
\sigma _t=\sigma _{el}+\sigma _{in},
\label{telin}
\end{equation}
i.e., that the total probability of all processes is equal to one. 

That allows to estimate the real and imaginary parts separately just in forward 
direction $t=0$ after the values of $\rho _0$ and $\sigma _t$ are measured. 

The real and imaginary parts of the amplitude are, in general, related at 
any $t$ by the dispersion relations as parts of a single analytic function. 
This approach was used for predictions of the energy behavior of $\rho _0$ 
and was only successful at the qualitative level because its accuracy 
depends on extrapolations of ${\rm Im}f(s,0)$ and, consequently, of 
$\sigma _t$ to higher energies.
The similar treatment at the arbitrary values of $t$ asks for some additional 
assumptions, and the conclusions strongly depend on them.

The theoretical approaches differ in ascribing different roles for the 
relative contributions of real and imaginary parts at $t\neq 0$. 
Unfortunately, the available tools are rather moderate and can not 
exploit the power of QCD at full strength. Elastic scattering 
implies that the same hadrons are observed in the final state. It means that
the partons inside them acted collectively while the QCD methods are
applicable to incoherent interactions of individual partons at high 
transferred momenta. Therefore, the phenomenological models and some insights 
from the unitarity relation are mostly used.  

We also know from experiment the energy behavior of real and imaginary parts of 
the amplitude (or their ratio $\rho (s,0)=\rho _0$) in forward direction $t=0$.
At high energies this ratio is rather small. For proton-proton scattering, it is 
negative at lower energies (reaching the values about -0.3), becomes equal to 
zero at energies about 100 GeV, exhibits a positive maximum and diminishes 
to about 0.1 at LHC energy 7 TeV. Most phenomenological models aim to fit 
experimental data about the differential cross section and the ratio 
$\rho (s,0)$ in a wide energy range and to predict them at higher energies. 
Up to now, we can not claim that the desired aim has been achieved as seen, 
in particular, in Fig. 1(right) where the failure of predictions of five 
theoretical models at 7 TeV is shown even though all of them were quite
successful at lower energies. 

This situation is described in more details in 
the review paper \cite{ufnel}. I will not repeat it here.
Instead, I concentrate on the described above two important problems. 

\section{The geometry of the interaction region}

The structure of protons is one of the main problems in particle physics.
It is well known \cite{kmwg} that up to now there exists the $\approx 7\sigma $
disagreement between the proton charge radius determined from muonic hydrogen 
and from electron-proton systems: atomic hydrogen and ep elastic scattering. 
In deep inelastic  electron-proton collisions the partonic structure of
protons is successfully studied. The point-like
nature is ascribed to the colliding electron. Therefore, the interaction region 
is defined by the proton size. Its size and opacity (or darkness) are determined 
by the Fourier-image of the generalized
parton distribution functions of protons depending on the total energy and the 
virtuality of the exchanged photon measured in experiment. Both the size and 
the opacity evolve with energy because the parton content of the proton evolves.

In proton-proton as well as in proton-nucleus and nucleus-nucleus
collisions, both objects possess some complicated internal 
structure. The partons of one of them can interact with many partons from 
another one distributed somehow within some space volume. Moreover, there
can be coherent interactions of some groups of partons. Therefore, it is hard 
to disentangle the individual contributions. The correlation femtoscopy, using 
its correspondence to the well known in astrophysics Hanbury-Brown and Twiss 
intensity interferometry, is 
widely applied for studies of the space-time structure in inelastic processes. 
The correlations between the momenta of newly created particles (mostly,
pions) reveal the space structure of the interaction region. This technique is 
especially successful in applications to nuclei but meets some problems 
\cite{sish} for smaller objects like protons connected with the Heisenberg
uncertainty relation. The uncertainty limit is about 1 fm for the current 
high energy experiments. The coherence of individual sources should be
taken into account for such systems.

Here, we show that it is possible to study the space structure of the 
interaction region of colliding protons even at smaller distances using 
the information about their elastic scattering. The transverse size of
this region will be discussed. We do not consider the longitudinal and time
sizes because they are strongly related to the model dependent suppositions
of the partonic structure of protons (the relative contributions of partons with 
definite shares of the longitudinal momentum). In fact, the role of the 
generalized parton distribution functions integrated over the longitudinal
momenta is to be studied. Experimental results about properties of the 
diffraction cone, which automatically account for nonperturbative dynamics of
the process and coherence of unknown internal sources, define main features of 
the transverse structure. The parameters obtained from experimental data about 
elastic processes are directly related to such properties of this region as 
its transverse size and opacity (or blackness). Their energy dependence 
determines its evolution with collision energy. 

To define the geometry of the collision we must express all characteristics
presented by the angle $\theta $ and the transferred momentum $t$
in terms of the transverse distance between the centers of the colliding
protons called the impact parameter $b$. It is easily done by the 
Fourier -- Bessel transform of the amplitude $f$ which retranslates the
momentum data to the transverse space features and is written as
\begin{equation}
i\Gamma (s,b)=\frac {1}{2\sqrt {\pi }}\int _0^{\infty}d\vert t\vert f(s,t)
J_0(b\sqrt {\vert t\vert }).
\label{gamm}
\end{equation}
The unitarity condition in the $b$-representation reads
\begin{equation}
G(s,b)=2{\rm Re}\Gamma (s,b)-\vert \Gamma (s,b)\vert ^2.
\label{unit1}
\end{equation}
The left-hand side (the overlap function in $b$-representation) describes the 
transverse impact-parameter profile of inelastic collisions of protons. It is 
just the Fourier -- Bessel transform of the overlap function $g$. It satisfies 
the inequalities $0\leq G(s,b)\leq 1$ and determines how absorptive is the 
interaction region depending on the impact parameter (with $G=1$ for the full 
absorption and $G=0$ for the complete transparency). The profile of elastic 
processes is determined by the subtrahend in Eq. (\ref{unit1}). If $G(s,b)$ is
integrated over the impact parameter, it leads to the cross section of
inelastic processes. The terms on the right-hand side would produce the total
cross section and the elastic cross section, correspondingly, as it should be
according to Eq. (\ref{telin}). The overlap 
function is often shown in relation with the opacity (or the eikonal phase) 
$\Omega (s,b)$ such that $G(s,b)=1-\exp (-\Omega (s,b))$. Thus, the full
absorption corresponds to $\Omega =\infty $ and the complete transparency to
$\Omega =0$. 

Even though the impact parameter can not be directly measured, the geometric
picture is instructive and closely related to such experimentally found
characteristics as the ratio of the diffraction cone slope to the total cross 
section that provides immediate guides to its energy evolution.
The impact parameter profiles of elastic and inelastic hadron collisions are 
derived as Fourier - Bessel transforms of the measurable data. They help us 
visualize the geometrical picture of partonic interactions indicating their 
space extension and the intensity. Our intuitive guesses about the space-time 
development of these processes can be corrected in this way.

The diffraction cone contributes mostly to the Fourier - Bessel transform of the
amplitude. Using the above formulae, one can write the dimensionless $\Gamma $ as
\begin{equation}
i\Gamma (s,b)=\frac {\sigma _t}{8\pi }\int _0^{\infty}d\vert t\vert 
\exp (-B\vert t\vert /2 )(i+\rho (s,t))J_0(b\sqrt {\vert t\vert }).
\label{gam2}
\end{equation}
Here, the diffraction cone approximation (\ref{expB}) is inserted. Herefrom, 
one calculates
\begin{equation}
{\rm Re}\Gamma (s,b)=\frac {1}{Z}\exp (-\frac {b^2}{2B}),
\label{rega}
\end{equation}
where we introduce the dimensionless ratio of the cone slope (or the elastic
cross section) to the total cross section
\begin{equation}
Z=\frac {4\pi B}{\sigma _t}=\frac{\sigma _t}{4\sigma _{el}}.
\label{ze}
\end{equation}
This dependence on the impact parameter was used, in particular, in \cite{fsw}.
Possible small deviations from the exponential behavior (see, e.g., \cite{anti, zrt})
inside the cone do not practically change the value of the integral contribution.
The differential cross section is quite small outside the diffraction peak
and also does not influence the impact parameter profile $G$. Therefore, our
task happens to be practically independent of the second problem.  

As was mentioned, the ratio $\rho (s,t)$ is very small at
$t=0$ and, at the beginning, we neglect it and get
\begin{equation}
G(s,b)= \frac {2}{Z}\exp (-\frac {b^2}{2B})-\frac {1}{Z^2}
\exp (-\frac {b^2}{B}).
\label{ge}
\end{equation}
For central collisions with $b=0$ it gives
\begin{equation}
G(s,b=0)= \frac {2Z-1}{Z^2}.
\label{gZ}       
\end{equation}
This formula is very important because it follows herefrom that
the darkness at the very center is fully determined by the parameter $Z$, 
i.e. by the ratio of experimentally measured characteristics - the width of the 
diffraction cone $B$ (or $\sigma _{el}$) to the total cross section.
Their energy evolution defines the evolution of the absorption value.
The interaction region becomes completely absorptive $G(s,0)=1$ in the center 
only at $Z=1$ and the absorption diminishes for other values of $Z$. 

In the Table, we 
show the energy evolution of $Z$ and $G(s,0)$ for $pp$ and $p\bar p$ 
scattering as calculated from experimental data about the total cross section
and the diffraction cone slope at corresponding energies.
\medskip
\begin{table}
\medskip
Table.  $\;\;$ The energy behavior of $Z$ and $G(s,0)$.
\medskip

    \begin{tabular}{|l|l|l|l|l|l|l|l|l|l|l|l}
        \hline
$\sqrt s$, GeV&2.70&4.11&4.74&7.62&13.8&62.5&546&1800&7000\\ \hline
$Z$           &0.64&1.02&1.09&1.34&1.45&1.50&1.20&1.08&1.00 \\  
$G(s,0)$     &0.68&1.00&0.993&0.94&0.904&0.89&0.97&0.995&1.00 \\  \hline
   
\end{tabular}
\end{table}
Let us point out that starting from ISR energies the value of $Z$ decreases
systematically and at LHC energies becomes equal to 1 within the accuracy of
measurements of $B$ and $\sigma _t$.

The impact parameter distribution of $G(s,b)$ (\ref{ge}) has the maximum 
at $b_m^2=-2B\ln Z$ with full absorption $G(b_m)=1$. 
Its position depends both on $B$ and $Z$. 

Note, that, for $Z>1$, one gets 
the incomplete absorption $G(s,b)<1$ at any physical $b\geq 0$ with the largest 
value reached at $b=0$ 
because the maximum appears at non-physical values of $b<0$. The disk is
semi-transparent. 

At $Z=1$, the maximum is positioned exactly at $b=0$, and the absorption is
absolutely strong there $G(s,0)=1$. The disk center becomes impenetrable 
(black). 

At $Z<1$, the maximum shifts to positive physical impact parameters. The dip 
is formed at the center that leads to the concave shape of the inelastic 
interaction region approaching the toroid shape. It becomes deeper at 
smaller $Z$. The limiting value $Z=0.5$ leading to the complete transparency 
at the center $b=0$ is considered in more details below.

The maximum absorption in central collisions $G(s,0)=1$ is reached at the 
critical point $Z=1$ which is the case at the LHC energy $\sqrt s=7$ TeV as
seen from the Table. Therefore it is considered first. 
Moreover, the strongly absorptive core of the interaction region grows in size
as we see from expansion of Eq. (\ref{ge}) at small impact parameters:
\begin{equation}
G(s,b)= \frac {1}{Z^2}[2Z-1-\frac {b^2}{B}(Z-1)-\frac {b^4}{4B^2}(2-Z)].
\label{gb}
\end{equation}
\begin{figure}
\includegraphics[width=\textwidth, height=6.2cm]{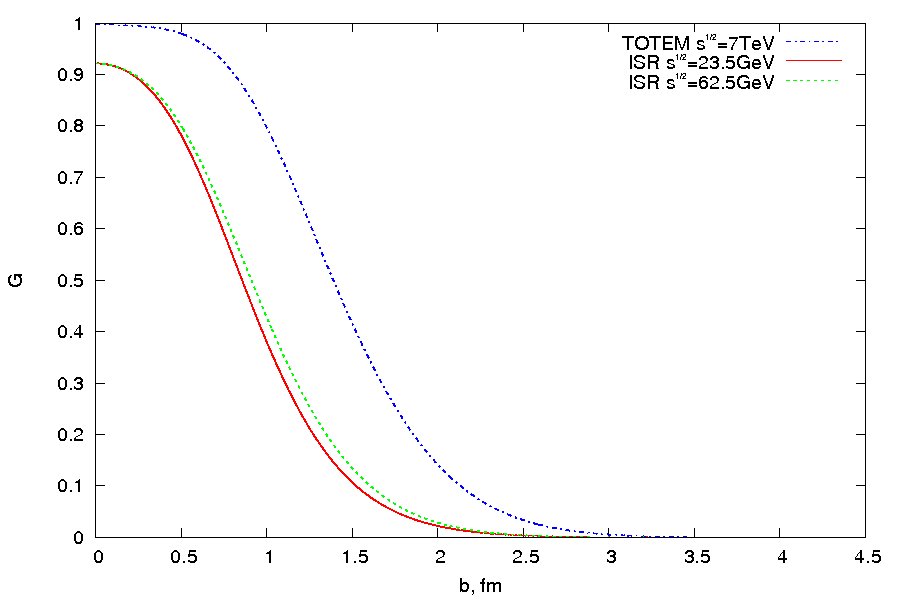}

Fig. 2. The overlap function $G(s,b)$ at 7 TeV (upper curve)
\cite{dnec} compared to those at ISR energies 23.5 GeV and 62.5 GeV
(all of them are computed by using the fit of experimental data according to 
the phenomenological model \cite{amal}). 
\end{figure}
The second term proportional to $b^2$ vanishes at $Z=1$, and $G(b)$ develops a 
plateau which extends
to quite large values of $b$ about 0.4 - 0.5 fm. The plateau is very flat
because the third term starts to play any role at 7 TeV (where 
$B\approx 20$ GeV$^{-2}$) only at even larger values of $b$. The structure of 
the interaction region with a central core at energies 7 - 8 TeV is also 
supported (see Fig. 2, where it is compared with corresponding structures
at ISR energies) by direct computation \cite{dnec} using the 
experimental data of the TOTEM collaboration \cite{totem1, totem2} about
the differential cross section in the region of $\vert t\vert \leq 2.5$ GeV$^2$.

The results of analytical calculations according to (\ref{ge}) and the direct 
computation practically coincide
(see Fig. 1 in \cite{ads}). It was also shown in \cite{ads} that this 
two-component structure with the central black core and more transparent 
periphery is well fitted by the expression with the abrupt 
(Heaviside-like) change of the exponential. However, it is still pretty far 
from the black disk limit because the peripheral region at $b$ near 1 fm is
very active and shows strong increase compared to ISR energies \cite{dnec}.
This is demonstrated in Fig. 3 where the difference 
$\Delta G(b)=G(s_1,b)-G(s_2,b)$ between the overlap functions at different 
energies $s_1$ and $s_2$ is displayed. 

The lower plot in Fig. 3, obtained in
\cite{amal}, demonstrates that even within the quite narrow interval of ISR 
energies the role of peripheral interactions with the impact parameters about
1 fm increases in inelastic processes with energy increase. Even more
spectacular is the peripheral increase when one comes from ISR to LHC energies
as seen in the upper plot in Fig. 3. Moreover, the darkness of the central
core strongly increases in Figs 2 and 3 that becomes especially important
as we discuss in detail in the next section. 
\begin{figure}
\includegraphics[width=\textwidth, height=7cm]{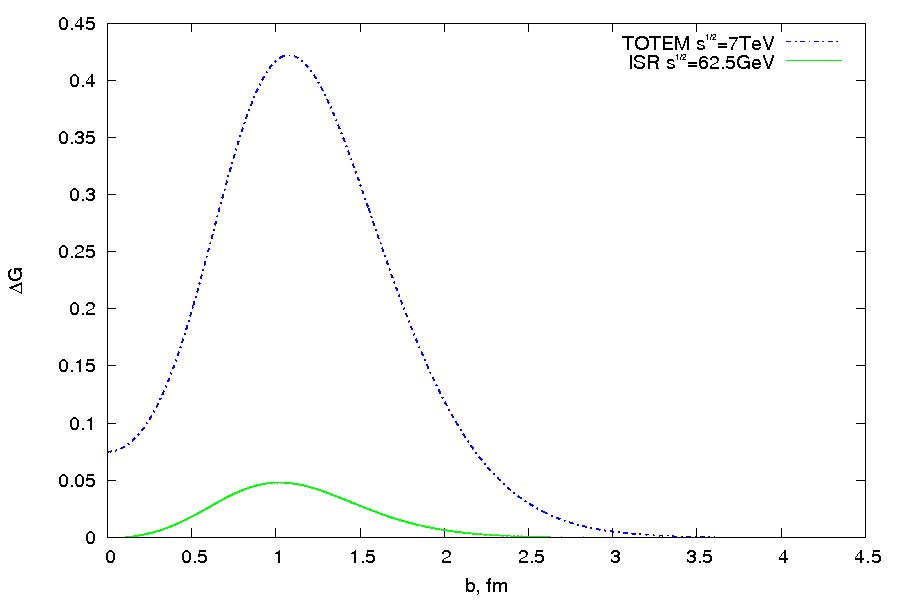}

Fig. 3. The difference between the overlap functions at different energies 
\cite{dnec, amal}. Dash-dotted curve is for 7~TeV and 23.5~GeV energies, 
solid curve is for 62.5~GeV and 23.5~GeV.

Conclusion: The parton density at the periphery strongly increases!
\end{figure}

It is interesting that the positivity of $G(s,b)$, i.e. of 
$\sigma _{inel}(s,b)$, imposes some limits on the relative role of $B$ and 
$\sigma _t$. Namely, it follows from Eq. (\ref{gZ}) that
\begin{equation}
2Z=\frac {8\pi B}{\sigma _t}=\frac{\sigma _t}{2\sigma _{el}}\geq 1.
\label{Bsig}
\end{equation}
This relation implies that the slope $B$ should increase with energy at
least as strong as the total cross section $\sigma _t$. 

This inequality is fulfilled at present and intermediate energies. If the value
of $Z$ will decrease at energies above 7 TeV, as one could expect from its
tendency shown in the Table, and approach $Z=0.5$ this inequality can be
saturated. We discuss first what happens in the region $0.5\leq Z\leq 1$.
The values $Z<0.5$ will be discussed at the end of this section.

It is usually stated that the equality 
$2Z=8\pi B/\sigma _t=\sigma _t/2\sigma _{el}=1$ corresponds to 
the black disk limit. Surely, the equality of elastic and inelastic cross 
sections is fulfilled $\sigma _{el}=\sigma _{in}=0.5\sigma _t$. However, 
beside this equality, the scattering on the black disk should result in the 
special non-exponential shape of the differential cross section of the type
\begin{equation}
\frac {d\sigma }{dt}\propto \frac {J_1^2(R\sqrt {\vert t\vert })}{\vert t\vert }.
\label{disk}
\end{equation}
It possesses a zero at $\vert t\vert \approx 3.67/B$ if the relation $R^2=4B$
is used. At the energy 7 TeV this zero should be placed at $\vert t\vert = 0.2$
GeV$^2$. That is not confirmed by experiment as well as the 
equality of cross sections of elastic and inelastic collisions. 

In principle, one can not exclude possible fast change of the regime of
the exponential decrease in the diffraction cone (\ref{expB}) by a new one.
The appearance of another, more steep exponent at the end of the diffraction 
cone in Fig. 1(left) at 7 TeV and the noticable shift of the dip position to 
smaller transferred momenta at LHC compared to lower energies could be the 
first signs of it.

Nevertheless, we continue to study the situation assuming that the exponential
regime valid up to LHC will persist at higher energies. One sees from 
Eq. (\ref{gZ}) that $G(s,b=0)=0$ at $Z=0.5$, i.e. the inelastic interaction 
region is completely transparent in central collisions. Surely, one should
not call it as a black disk. This paradox is 
resolved \cite{drjl} if we write the inelastic profile of the interaction region using
Eq. (\ref{ge}). At $Z=0.5$ it looks like
\begin{equation}
G(s,b)= 4[\exp (-\frac {b^2}{2B})-\exp (-\frac {b^2}{B})].
\label{0.5}
\end{equation}
We see that one must rename the black disk as
a black toroid (or a black ring if we consider its two-dimensional projection) 
with full absorption $G(s,b_m)=1$ at the 
\begin{figure}[hbtp]
\centerline{\includegraphics[width=\textwidth, height=9cm]{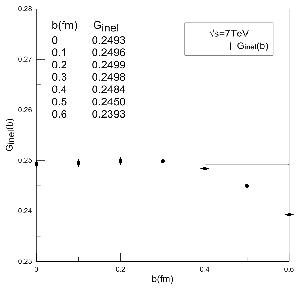}}

Fig. 4. The impact parameter dependence of the function $G_{inel}(b)=0.25 G(b)$
at 7 TeV \cite{mart}. It is obtained using the fit of experimental data 
according to the model \cite{amal}.
\end{figure}
impact parameter 
$b_m=R\sqrt {0.5\ln 2}\approx 0.59R$, complete transparency at $b=0$ and rather
large half-width about 0.7$R$. Thus, the evolution to values of $Z$ smaller than 
1 at higher energies (this can happen if the decreasing tendency of $Z$ with 
energy shown in the Table persists) would imply quite special transition from 
the critical two-scale regime at the LHC to the concave toroid-like (or ring-like 
in two dimensions) configurations of the interaction region if the exponential 
shape of the diffraction cone, described by Eq. (\ref{expB}), persists. 
 
It looks as if the protons penetrate through one another at central collisions, 
just scattering elastically, while peripheral collisions become responsible for 
inelastic processes. Elastic and inelastic profiles get equal only at $b=b_m$. 
Elastic one dominates at $b<b_m$ while inelastic one at periphery $b>b_m$.

Is the parton coherence inside each colliding proton
responsible for that? Can we observe
its effects similar to difference of light scattering in the water (coherence!)
and in the air (decoherence and fluctuations are in charge of the blue color of 
the sky!)? Elastic and inelastic profiles get equal only at $b=b_m$. Elastic 
one dominates at $b<b_m$ while inelastic one at periphery $b>b_m$.

Paradoxically enough, it pushes us back to the
early suggestions that inelastic processes are more peripheral (recall the
one-pion exchange model!) than elastic scattering which is a shadow of
inelastic interactions (more pions exchanged). This tendency is clearly seen
already at present energies as demonstrated in Fig. 3. We stress that the 
total region of proton interactions keeps the Gaussian shape
described by the first term in the right-hand side of Eq. (\ref{ge}).
The Gaussian shape is preserved for the elastic profile as well but with the
twice steeper exponent (see Eqs (\ref{0.5}), (\ref{ge})).

What concerns the longitudinal distances, it is commonly believed according to
the parton model that they are much larger than the transverse size, especially
for soft partons. Then the region of inelastic interactions would remind just 
a toroid, i.e. a tube, at the center of which only elastically scattered protons 
fly.

The plateau of $G(b)$ at small $b$ was confirmed in \cite{mart} as shown
in Fig. 4. However, an additional substructure at the level 10$^{-4}$ in 
a single point has been mentioned there. The function
$G_{inel}=0.25G$ at 7 TeV, plotted there, slightly decreases at $b=0$ compared 
to its values at $b$=0.1 -  0.3 fm (compare the numerical values shown in the 
Figure). This could be an indication that $Z$ becomes somewhat
less than 1 already at 7 TeV and the transition to the concave shape starts at 
this energy. At the same time, no decrease is seen in Fig. 2. This disagreement
is especially surprising because the same model \cite{amal} was used in 
papers \cite{dnec, mart} to fit experimental results at 7 TeV. However, we see 
that the excess at impact parameters 0.1 - 0.3 fm compared to the center 
pointed out in \cite{mart} reveals itself in the fourth digits only while
the error bars of the slope $B$ and the total cross section $\sigma _t$ are
larger by an order of magnitude. This excess is so small that it can be 
explained either by insufficient precision in determining the values of $B$ and 
$\sigma _t$ (and, consequently, of $Z$) and also by inaccuracy in account for 
different scales or by different procedures adopted in the papers
\cite{dnec, mart} for extrapolations to the ranges of transferred momenta 
where there are no experimental data yet.

Thus it seems too early to make any (even preliminary) statements. However, the 
comparison of the results of \cite{dnec} and \cite{mart} shows that we are in 
the critical regime of elastic scattering with $Z\approx 1$ at 7 TeV as was 
pointed out in \cite{drjl}. Let us stress that at good precision of experimental 
data the proposed approach allows to analyse the fine structure of the core of
the interaction region at the very small scales in distinction to the
less precise correlation methods. Therefore we
should pay special attention to evolution of the parameter $Z$ at higher
energy of 13 TeV which will become available soon. The especially precise
measurements of the diffraction cone slope $B$ and the total cross section
$\sigma _t$ would be very desirable.

Another consequence of Eq. (\ref{gZ}) follows from study of energy evolution of 
$G(s,0)$ shown in the Table. In connection with the torus-like concave structure, 
it is interesting to point out the value of $Z=0.64$ or $G(s,0)=0.68$ at 
$\sqrt s=2.70$ GeV and maximum 1 at $b_m^2=4B\ln 2$. However, at this rather 
low energy the whole analysis should be redone with account of the diffraction 
cone behavior, total cross section and the value of the ratio $\rho $. One also 
notices that, in 
the energy interval 4 GeV$<\sqrt s<8$ GeV, the values of $Z$ are slightly larger 
than 1 so that the values of $G(s,0)$ are smaller but very close to 1. These 
facts ask for further studies in the energy interval 2.7 GeV$<\sqrt s<8$ GeV 
especially in view of proposed experiments in Protvino. 
The dark core must be smaller at lower energies than at LHC because of smaller 
values of $B$. Moreover, the contribution due to the real part of the amplitude is larger 
at these energies as well as larger $\vert t\vert $ beyond the diffraction cone 
can be important. The dependence of $Z$ on energy shown in the Table looks
as if the interaction region at low energies becomes black at the center $b=0$ 
but at higher energies up to ISR loses this property trying to restore it at 
the LHC. 

In principle, the positivity of the inelastic cross section
\begin{equation}
\sigma _{in}=\frac {\pi B}{Z^2}(4Z-1)\geq 0
\label{posin}
\end{equation}
admits the value of $Z$ as small as 0.25 which asymptotically corresponds to 
$\sigma _{el}=\sigma _t$ and $\sigma _{in}=0$. The values of $Z<0.5$ lead
to negative values of $\sigma _{inel}(s,b)$, i.e. to negative Fourier - Bessel 
transforms 
of $g(p,\theta )$ in Eq. (\ref{gmm}). They are not forbidden if the relative 
phases of matrix elements of inelastic processes $M_i$ in Eq. (\ref{gmm}) 
interfere in such a way. Unfortunately, we have no knowledge about them.
This possibility was treated as another branch of the solution of the 
unitarity condition and named as antishadowing or refractive scattering in
\cite{trt} and as resonant disk modes in \cite{ann}. However, the value of $Z$
is close to 1 even at LHC energies, and this regime asking for $Z<0.5$ is,
surely, shifted to extremely high energies if it can be observed at all.
The approach to asymptotics is argued as following the logarithmic dependences
of cross sections $\sigma _t \propto \sigma _{el} \propto \ln ^2s$ and
$\sigma _{in} \propto \ln s$. The depletion of $G(s,0)$ in \cite{mart} was 
ascribed to this regime by mistake since the values of $Z$ are near 1 at 7 TeV
but not as small as 0.5.

\section{New tendencies of inelastic collisions}

The maximum absorption at central collisions at LHC energies must reveal itself 
in some special features of inelastic collisions in such a critical regime with
$Z\approx 1$.
The diffraction cone contributes mostly to $G(s,b)$. The large-$\vert t\vert $ 
elastic scattering can not serve as an effective trigger of the black core. One 
of the typical features of the high energy inelastic processes is the 
production of high energy jets, i.e. collimated groups of particles. The 
inelastic exclusive processes can be more effectively used for the analysis of 
the central black core. One needs such triggers which enhance its contribution. 
Following the suggestions of \cite{fsw, fsw1}, it becomes possible \cite{ads} 
to study the details of the central core using the experimental data of CMS 
collaboration at 7 TeV about inelastic collisions with high multiplicity 
triggered by the hadron jet production \cite{cms} as well as some 
other related data. Triggers (charged particles or jets) with large transverse
momenta are produced in central collisions. Therefore, the black plateau in the 
central part of the interaction region with $b<0.4  -  0.5$ fm should result in 
the corresponding plateau of the charged particle density in the transverse 
region $60^0<\vert \Delta \phi \vert <120^0$ defined as follows
\begin{equation}
\mu _{tr}=\frac {N_{ch}^{tr}}{\Delta \eta\Delta(\Delta \phi )},
\end{equation}
where $N_{ch}^{tr}$ is the charged particle multiplicity in the transverse 
region, $\Delta \eta $ is the pseudorapidity range studied,
$\Delta(\Delta \phi )$ is the azimuthal width of the transverse region.
This is really the case as shown in Fig. 5. 
\begin{figure}[hbtp]
\begin{center}
\includegraphics[ width=\textwidth, height=7cm]{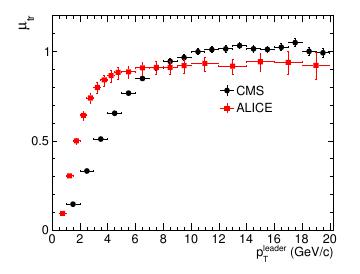}%

Fig. 5. Charged-particle density in the transverse region as a function of 
$p_{\rm T}$ of leading object \cite{ads} (CMS - charged-particle jet, 
ALICE - charged particle). CMS analyses  particles with $p_{\rm T}>$~0.5~GeV/{\it c} and
$|\eta|<$~2.4, \\ ALICE - with $p_{\rm T}>$~0.5~GeV/{\it c} and $|\eta|<$~0.8.
\label{M_tr}%
\end{center}
\end{figure}

Let us explain it. Starting from large transverse momenta of triggers on the
righthand side of Fig. 5 and going to the left, we somehow feel at the beginning
the central region from $b=0$ to the end of the plateau. Then the density of
accompanying particles in the transverse region should not change until we
approach the end of it. Only then the decrease of the distribution of the
accompanying particles should start. The difference in positions of such a 
decrease in the two plots is defined by the difference in the choice of the 
leading trigger used by the two collaborations. The flat dependence of 
$\mu _{tr}$ on $p_{\rm T}$ shows that activity in the transverse region is 
independent of hard process scale, provided that scale is hard enough that 
all proton-proton interactions are central.

Many other characteristics of such inelastic processes considered in 
\cite{ads} support this conclusion. Usage of very high multiplicity events in 
combination with jet properties is crucial. In particular, the most important
observation is that the significant reduction of jet rate at very high 
multiplicities compared to MC predictions asks for new inputs in the models.
Separating the core contribution with the help of these triggers, 
one comes to the important conclusion that the simple increase of the 
geometrical overlap area of the colliding protons does not account for
properties of jet production at very high multiplicities. It looks as if the 
parton (gluon) density must strongly increase in central collisions and
rare configurations (fluctuations) of the partonic structure of protons are 
involved. Up to now, the paper \cite{ads} is the first attempt to explain
the distinctive features of jets and underlying event properties at LHC
energies. The correlation studies of jets (see, e.g., \cite{acms}) can
be used for further femtoscopy of the fine structured system. Surely, some 
further proposals to learn the critical regime at 7 - 8 TeV will be proposed.

At the same time, implications of energy evolution of $Z$ (if it will be 
observed!) for inelastic processes are of great interest. The relative roles 
of elastic and inelastic cross sections will start approaching one another
in the case if values of $Z$ would diminish below 1 with increase of energy.
The mean multiplicity will, probably, decrease because of the more 
peripheral origin of newly created particles. The decreasing role of central 
interactions can lead to some changes in the shape of the multiplicity 
distributions (lower tails?) and diminished share of jets which will 
acquire new features. Jets will become produced at the periphery
in distinction to the situation described above. That would imply that they
will have to penetrate larger distances in transverse direction compared to
forward directions. It would give rise to their stronger depletion in the
transverse plane and, therefore, to the azimuthal asymmetries which were looked 
for in \cite{acms}. Surely, there will be found other criteria of transition to 
the concave shape of the interaction region of inelastic processes.

Hardly, we will be able to reach the regime with extremely small values
of $Z<0.5$ when inelastic processes would play negigibly small role 
compared to elastic ones.

\section{Elastic scattering outside the diffraction cone}

This problem attracts special attention since 60th when the experimental data 
about the differential cross section at high energies and rather large 
transferred momenta appeared for the first 
time. It was observed that the exponential $t$-regime in the diffraction cone
is replaced by the exponential $\sqrt t$-regime at larger angles. 
The latter region of angles was called the Orear region in the name of its
investigator.  

The unitarity condition happened to be very successful in that region of angles 
as well. Theoretically, it can be approached considering the unitarity condition 
(\ref{unit}) directly in $s,t$-variables without using the Fourier - Bessel
transform as it was done at small angles. It was shown a long time ago 
\cite{anddre1, anddre2} that
the imaginary part of the amplitude $f$ outside the diffraction cone
can be derived from the general unitarity condition (\ref{unit}) 
which is reduced there to the inhomogeneous linear integral equation
\begin{equation}
{\rm Im}f(p,\theta )=\frac {p\sigma _t}{4\pi \sqrt {2\pi B}}\int _{-\infty }
^{+\infty }d\theta _1 \exp (-Bp^2(\theta -\theta _1)^2/2) r_{\rho }
{\rm Im}f(p,\theta _1)+g(p,\theta ),
\label{linear}
\end{equation}
where $r_{\rho }=1+\rho (s,0)\rho (s,\theta _1) $. 
This reduction becomes possible because the contribution from asymmetrical 
configuration of scattering angles in the first term of Eq. (\ref{unit})
dominates due to the steep Gaussian fall-off inside the diffraction cone. 
Because of the sharp fall off of the amplitude with angle, the 
principal contribution to the integral arises from a narrow region around the
line $\theta _1 +\theta _2 \approx \theta $. Therefore one of the amplitudes
should be inserted at small angles within the cone as a Gaussian while another 
one is kept at angles outside it.

It can be solved analytically (for more details see \cite{anddre1, anddre2})
with two assumptions that the role of the overlap function $g(p,\theta )$ is 
negligible outside the diffraction cone and the function $r_{\rho }$ may be
approximated by a constant, i.e. $\rho (\theta _1)=\rho _l$=const.
                           
Let us assume that the overlap function is negligible at these transferred
momenta\footnote{The assumption of smallness of the overlap function outside 
the diffraction cone is appealing intuitively. The point is that particles 
newly created in high energy inelastic processes move mainly inside narrow 
angular cones along the directions of the primary hadrons. Therefore, the 
geometrical overlap of two narrow cones, whose axes are turned by the
comparatively large angle $\theta $, is negligibly small. Moreover, it was 
shown to be valid 
\cite{ands, dnec1} by direct computation of the overlap function from 
experimental data in a wide energy interval up to LHC energies.}. Then, 
the eigensolution of the homogeneous linear integral equation is
\begin{equation}
{\rm Im} f(p,\theta )=C_0\exp \left (-\sqrt 
{2B\ln \frac {Z}{r_{\rho }}}p\theta \right )+\sum _{n=1}^{\infty }C_n
\exp (-({\rm Re }b_n)p\theta ) \cos (\vert {\rm Im }b_n\vert p\theta-\phi _n)
\label{solut}
\end{equation}
with
\begin{equation}
b_n\approx \sqrt {2\pi B\vert n\vert}(1+i{\rm sign }n) \;\;\;\;\;\;\; n=\pm 1, \pm 2, ...
\label{bn}
\end{equation}
This expression contains the exponentially decreasing with $\theta $ (or 
$\sqrt {\vert t \vert }$) term (Orear regime!) with imposed on it oscillations 
much stronger damped by their own exponential factors $b_n$ compared to
the first term. The critical role in the exponent of the first leading term 
which determines the rate of decrease of the differential cross section is
again played by the widely used above parameter $Z$ (\ref{ze}).
The oscillating terms become pronounced only at smaller angles and reveal 
themselves as the dip in the vicinity of the diffraction cone. 
The elastic scattering differential cross section outside 
the diffraction cone (in the Orear regime region) is
\begin{eqnarray}
\frac {d\sigma }{p_1dt}&= &\left (   e^{-\sqrt 
{2B\vert t\vert \ln \frac {Z}{r_{\rho }}}}
\right.
\nonumber \\
&+&\left.
 p_2e^{-\sqrt {2\pi B\vert t\vert}} \cos (\sqrt {2\pi B\vert t\vert }-\phi)
\right )^2     
\label{fit}
\end{eqnarray}
with the parameters $p_1$ and $p_2$ tightly related to $C_0$ and $C_1$.
Namely this formula was used in Refs \cite{adg, dnec1} for fits of experimental 
data about differential cross sections in a wide energy range.
The ratio $\rho $ was approximated by its average values in and outside the
diffraction cone so that $r_{\rho }=1+\rho _0\rho _l$ where $\rho _l$ is 
treated as the average value of $\rho $ in the Orear region. 

The fits at 
comparatively low energies \cite{adg} are consistent with $r_{\rho }\approx 1$,
i.e., with small values of $\rho _l$ close to zero. When $Z=1$, as it happens
at 7 TeV (see the Table above), the exponent of the main first term is very
sensitive to the value of $\rho _l$ outside the diffraction cone. For the 
first time, that allowed to estimate the value of $\rho $ in the Orear region 
at 7 TeV \cite{dnec1}. The great surprise of the fit of TOTEM data was necessity 
to use the large (in modulus) negative value of $\rho _l\approx -2.1$ if 
$\rho _0=0.14$ (as it was at ISR energies). 
Otherwise, the slope of the first term of Eq. (\ref{solut}) in the Orear region 
would be predicted to be equal to zero (constancy!) if $Z=1$ and $r_{\rho }=1$. 
It becomes larger in modulus $\rho _l\approx -3$ if the TOTEM value 
$\rho _0=0.1$ obtained at 7 TeV is used. Moreover, these values of $\rho _l$ 
can be considered as upper limits
because the effective value of $\rho _0$ inside the diffraction cone 
can be even smaller in view of its widely discussed zero there. 
No models have yet explained this finding. Further progress in solving
the unitarity equation with proper dependence of $\rho $ in and outside the
diffraction cone is needed.

The slope of the differential cross section in the Orear region becomes
a very sensitive indicator of the mutual behavior of $Z$ and $\rho _l$.
Possible decrease of the value of $Z$ with increasing energy and transition
to the torus regime, discussed above, would ask for further evolution
of the ratio $\rho $ to increasing in modulus negative values.

The predictive power of the solution (\ref{solut}) lies in its exponential
behavior in $\sqrt {\vert t\vert }$ with quite definite analytically
calculable exponent and imposed on it oscillations. Unfortunately, one 
can not definitely state where
its bounds on the $t$-axis are and has to rely on accuracy of fits within
some range of $t$. Nevertheless, as was mentioned above,
some important estimates of the value of
$\rho $ in the Orear region have been done. Another shortcoming of the 
of the solution (\ref{solut}) is its ignorance of the energy dependence.
Therefore, no predictions are made concerning higher energies.
Not only the normalization coefficients $C_0, C_n$ are unknown but the
exponent of the leading term suffers also from unpredictable energy
behavior of $B, \sigma _t$ and $r_{\rho }$. Only with results on $B, \sigma _t$
obtained from experiment at higher energies it is possible to estimate the 
ratio $\rho $ in the Orear region assuming it to be constant in there.

In parallel, there exists a variety of phenomenological models with numerous 
adjustable parameters which were proposed in attempts to describe experimental 
data. The theoretical arguments in favour of them and their main features 
are reviewed in some detail in \cite{ufnel}. Most of them describe quite well 
(albeit with some precaution) the behavior of the differential cross section 
in the diffraction cone. Therefore, if applied to the problem, they would 
reproduce the main features of 
the shape of the interaction region discussed in the previous section.
The imaginary part of the amplitude dominates in there. At the same time,
they failed in their predictions at 7 TeV in the Orear region as shown in
Fig. 1(right). The main problem lies in their inability to predict the energy
dependence of adjustable parameters. Only some qualitative guesses can be used.

Surely, it is much easier to use such guesses and
to fit these parameters by the newly available data.
That is what was successfully done a'posteriori, for example, in papers
\cite{dnec, mart, kfk}. First two of them used the model \cite{amal}
proposed a long ago, while a completely new form of the amplitude was 
analyzed in \cite{kfk}. 

Independently of success or failure of these models in describing the
Orear region, all of them have a common heuristic feature. To explain such 
prominent characteristics of the differential cross section as the dip
(see Fig. 1(right)), they have to assume that it originates at the 
transferred momentum $t$ where the imaginary part of the amplitude 
(dominating until then within the diffraction cone!) becomes equal to zero. The 
real part contributes in the dip to the differential cross section. The models
differ in the numbers of zeros in real and imaginary parts and in their
positions except the definitely fixed $t$-value at the dip.  

\begin{figure}[hbtp]
\begin{center}
\includegraphics[width=0.7\textwidth, height=6cm]{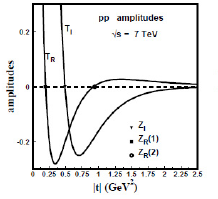}%

Fig. 6. Real $T_R={\rm Re} f$ and imaginary $T_I={\rm Im} f$ parts of the 
proton-proton amplitude at 7 TeV according to a particular phenomenological 
model \cite{kfk}.
\label{reim}%
\end{center}
\end{figure}
As an example, I 
show in Fig. 6 the corresponding graphs obtained for the model \cite{kfk}.
The zero of the imaginary part at the dip position is marked as $Z_I$.
It is its only zero. The real part possesses two zeros. One of them, $Z_R(1)$, 
lies within the diffraction cone. This zero is typical for many models and 
was somehow envisaged in papers \cite{mar1, mar2} even though without definite 
predictions about its position. Its appearence leads to 
necessity to diminish the theoretical estimates of the average value of 
$\rho $ inside the cone which enters in the soluion (\ref{solut}) as $\rho _0$ 
but should be treated, strictly speaking, as the average value of $\rho $ 
inside the cone. In its turn, that would lead to larger
(in modulus) values of $\rho _l$ as discussed above. The peculiar feature of
the model \cite{kfk} is that the real part crosses the abscissa axis and 
possesses the second zero $Z_R(2)$. Therefore, the ratio $\rho $ becomes 
negative in the Orear region albeit not large (in modulus) enough to
correspond to estimates done in the unitarity approach. Anyway, there is
some correspondence between them at the qualitative level. Other models have 
usually single zeros of real and imaginary parts and can not get the
negative values of $\rho $ in the Orear region. Probably, this is the
origin of their failure in predicting the LHC results. 

It is worthwhile here to stress that no zeros of the imaginary part are
predicted in the unitarity approach. The dip is explained \cite{dnec1}
as the contribution of the oscillatory terms in (\ref{solut}) whose role 
increases at smaller transferred momenta. At the same time, one should
critically remark that fits of the differential cross section in the Orear
region with the help of the imaginary part of the amplitude only becomes
selfcontradictory after the conclusion about the large value of $\rho _l$
in there is obtained. No consensus has been reached yet
between these two approaches. In general, one can state that until now the
$t$-dependence of real and imaginary parts is not understood theoretically.  

\section{Conclusions}

The usage of the indubitable general principle - the unitarity condition in
combination with experimental results about the elastic scattering in the
diffraction cone makes it possible to reveal the space image of the
interaction region of protons and its evolution with energy as well as
to estimate for the first time the average value of the real part of the
elastic scattering amplitude beyond the diffraction cone.

The behavior of the real and imaginary parts of the elastic scattering 
amplitude as functions of energy $s$ and transferred momentum $t$ completely 
defines the properties of this process and, in some way, influences the 
properties of inelastic processes through the unitarity requirement. Our 
knowledge of this behavior is still quite limited. 
QCD methods do not work. The general QCD statements and phenomenological models
are usually considered. 

It is shown that the geometrical space shape of the interaction region of 
protons is mainly determined by their elastic scattering at small angles.
The absorption at its center is determined by a single energy-dependent
parameter $Z$. The region of full absorption extends to quite large impact 
parameters up to 0.5 fm if $Z$ tends to 1. This happens at the LHC energy 
$\sqrt s=7$ TeV where the critical two-scale (the black central core and 
the more transparent peripheral region) structure of the interaction region of 
protons becomes well pronounced. The sharp separation of these two regions 
leads to special consequences both for elastic and inelastic processes.
The behavior of the parameter $Z$ at higher energies is especially important 
for the evolution of the geometry of the interaction region. The assumption
about its further decrease at higher energies results in drastic change of the 
geometry predicting the tendency for evolution to the completely unexpected 
toroid (tube)-like (or ring-like in two dimensions) configuration, where
the core becomes absolutely penetrable, and the complete absorption region is
shifted to some finite impact parameter. 
                                                               
The value of $Z=1$ attained at the LHC energies is also crucial for the 
behavior of the elastic scattering differential cross section 
outside the diffraction cone. In this case, its slope there becomes fully 
defined by the ratio of the real part of the amplitude to its imaginary part 
which is yet unknown at LHC energies in this range of the 
transferred momenta. No ways to its direct measurement is foreseen nowadays.
Therefore, it is very important that the analysis of experimental data at 7 TeV 
about the slope of 
the differential cross section inside the Orear region with the help of the
unitarity condition provides the estimate of its average value and
reveals that this ratio is negative and surprisingly large in modulus. The 
predictions of phenomenological models are contradictory in this region of 
transferred momenta. In general, these models should be checked for their 
selfconsistency by calculation of the overlap functions $g(p,\theta )$ for 
each of them. This is possible because the integral in the unitarity relation 
(\ref{unit}) can be computed with both real and imaginary parts of the amplitude 
for a particular model known. The measurement of the rate of decrease of the
differential cross section in the Orear region becomes very important at
higher energies because it happens to be very sensitive to the mutual 
behavior of $Z$ and $\rho _l$ with increase of energy.

Thus, the unitarity condition provides us many inspiring guides about
hadron interactions which should be taken into account by other approaches.

I am grateful for support by the RFBR grants 12-02-91504-CERN-a,
14-02-00099  and the RAS-CERN program.


\begin{thebibliography}{99}
\bibitem{totem1}
Antchev G et al. (TOTEM Collab.) {\it Europhys. Lett.} {\bf 95} 41001 (2011) 
\bibitem{totem2}
Antchev G et al. (TOTEM Collab.) {\it Europhys. Lett.} {\bf 96} 21002 (2011)
\bibitem{anti}
Antipov Yu M et al. {\it Phys. Rev. Lett.} {\bf 35} 1406 (1975)
\bibitem{zrt}
Zotov N P, Rusakov S V, Tsarev V A {\it Elem Part and Nucl} {\bf 11} 1160 (1980)
\bibitem{anddre1}
Andreev I V, Dremin I M {\it ZhETF Pis'ma} {\bf 6} 810 (1967) 
\bibitem{ufnel}
Dremin I M {\it Physics-Uspekhi} {\bf 56} 3 (2013)
\bibitem{hove}
Van Hove L {\it Nuovo Cimento} {\bf 28} 798 (1963)
\bibitem{kmwg}
Kraus E, Mesick K E, White A, Gilman R, Strauch S, arXiv:1405.4735 
\bibitem{sish}
Sinyukov Yu M, Shapoval V M {\it Phys. Rev. D} {\bf 87} 094024 (2013) 
\bibitem{fsw}
Frankfurt L, Strikman M, Weiss C {\it Phys. Rev. D} {\bf 83} 054012 (2004) 
\bibitem{dnec}
Dremin I M, Nechitailo V A {\it Nucl. Phys. A} {\bf 916} 241 (2013)
\bibitem{amal}
Amaldi U, Schubert K R {\it Nucl. Phys. B} {\bf 166} 301 (1980)
\bibitem{ads}
Azarkin M Yu, Dremin I M, Strikman M {\it Phys. Lett. B} {\bf 735} 244 (2014) 
\bibitem{drjl}
Dremin I M {\it JETP Lett.} {\bf 99} 243 (2014)
\bibitem{mart}
Alkin A, Martynov E, Kovalenko O, Troshin S M {\it Phys. Rev. D} {\bf 89}
091501(R) (2014)
\bibitem{trt}
Troshin S M, Tyurin N E {\it Phys. Lett. B} {\bf 316} 175 (1993)
\bibitem{ann}
Anisovich V V, Nikonov V A, Nyiri J, arXiv:1408.0692
\bibitem{fsw1}
Frankfurt L, Strikman M, Weiss C {\it Phys. Rev. D} {\bf 69} 114010 (2004) 
\bibitem{cms}
CMS Collaboration {\it Eur. Phys. J. C} {\bf 73} 2674 (2013)
\bibitem{acms}
CMS Collaboration PAS FSQ-13-005 (2013)
\bibitem{anddre2}
Andreev I V, Dremin I M {\it Sov. J. Nucl. Phys.} {\bf 8} 473 (1969)
\bibitem{ands}
Andreev I V, Dremin I M, Steinberg D N {\it Sov. J. Nucl. Phys.} {\bf 11} 261 
(1970)
\bibitem{dnec1}
Dremin I M, Nechitailo V A {\it Phys. Rev. D} {\bf 85} 074009 (2012)
\bibitem{adg}
Andreev I V, Dremin I M, Gramenitskii I M {\it Nucl. Phys.} {\bf 10} 137 (1969)
\bibitem {kfk}
Kohara A K, Ferreira E, Kodama T {\it Eur. Phys. J. C} {\bf 73} 2326 (2013);
arXiv:1408.1599
\bibitem{mar1}
Martin A {\it Lett. Nuovo Cim.} {\bf 7} 811 (1973)
\bibitem{mar2}
Martin A {\it Phys. Lett. B} {\bf 404} 137 (1997)

\end{thebibliography}
\end{document}